\documentclass[fleqn,12pt]{article}
\usepackage{amsmath,amsfonts,epsfig}

\usepackage{bbold}
\numberwithin{equation}{section}

\newcommand{\1}{\mathbb{1}}

\def\be{\begin{equation}}       \def\eq{\begin{equation}}
\def\ee{\label{abc}  \end{equation}}         \def\eqe{\label{abc}  \end{equation}}

\def\bea{\begin{eqnarray}}      \def\eqa{\begin{eqnarray}}
\def\ena{\end{eqnarray}}        \def\eea{\end{eqnarray}}
                                \def\eqae{\end{eqnarray}}

\def\a{\alpha}
\def\b{\beta}

\def\d{\delta}
\def\e{\epsilon}           
\def\f{\phi}               
\def\g{\gamma}

\def\i{\iota}

\def\k{\kappa}                    
\def\l{\lambda}
\def\m{\mu}
\def\n{\nu}
  \def\w{\omega}
\def\p{\pi}                
  \def\th{\theta}                  
\def\r{\rho}                                     
\def\s{\sigma}                                   
\def\t{\tau}

\def\x{\xi}

\def\F{\Phi}
\def\G{\Gamma}

\def\L{\Lambda}
  \def\W{\Omega}
\def\P{\Pi}

\def\qwe{\Theta}

\def\ca{{\cal A}}

\def\cf{{\cal F}}
\def\cg{{\cal G}}

\def\cj{{\cal J}}

\def\cl{{\cal L}}
\def\cm{{\cal H}}

\def\cp{{\cal P}}

\def\cs{{\cal S}}

\def\cv{{\cal V}}

\def\bop#1{\setbox0=\hbox{$#1M$}\mkern1.5mu
        \vbox{\hrule height0pt depth.04\ht0
        \hbox{\vrule width.04\ht0 height.9\ht0 \kern.9\ht0
        \vrule width.04\ht0}\hrule height.04\ht0}\mkern1.5mu}
\def\pa{\partial}                              
\def\we{\wedge}                                         
\def\>{\rangle} 

\def\<{\langle} 

\def\Tilde#1{\widetilde{#1}}                   

\def\to{\rightarrow}

\def\gmn{g_{\m \n}}
\def\bmn{b_{\m\n}}
\def\hmnp{H_{\m\n\r}}

\def\pa{\partial}

\def\half{{1 \over 2}}

\def\ha{\frac12}                               

\def\IZ{\relax\ifmmode\mathchoice
{\hbox{\cmss Z\kern-.4em Z}}{\hbox{\cmss Z\kern-.4em Z}}
{\lower.9pt\hbox{\cmsss Z\kern-.4em Z}} {\lower1.2pt\hbox{\cmsss
Z\kern-.4em Z}}\else{\cmss Z\kern-.4em }\fi}
\def\IC{\relax\hbox{$\inbar\kern-.3em{\rm C}$}}
\def\IR{\relax{\rm I\kern-.18em R}}

\def\bx{{\bf X}}

\ifx\ltimes\undefined
\newcommand{\ltimes}{{\kern3pt\hbox{\vrule width 0.4pt height 5.30pt
depth .0pt}\kern-1.76pt\times\kern1pt}} \fi

\def\be{\begin{equation}}
\def\ee{\label{abc}  \end{equation}}
\def\ba{\begin{eqnarray}}
\def\ea{\end{eqnarray}}
\def\bq{\begin{quote}}
\def\eq{\end{quote}}

\def\part{\partial}

\def\beq{\begin{equation}}
\def\eeq{\label{abc}  \end{equation}}
\def\beqa{\begin{eqnarray}}
\def\eeqa{\end{eqnarray}}

\def\we{ \wedge}

\def\ti{\Tilde}

\def\Z {\mathbb{Z}}
\def\R {\mathbb{R}}

\def\XX {\mathbb{X}}

\def\cx{{\XX}}

\begin{document}
\thispagestyle{empty}
\begin{flushright}
hep-th/0605149\\
 Imperial/TP/06/CH/02 \\
 \end{flushright}\vskip 0.8cm\begin{center}
\LARGE{\bf   Doubled Geometry and T-Folds}
\end{center}
\vskip 0.6in

\begin{center}{\large C M  Hull }
\vskip 0.6cm{ Theoretical Physics Group,  Blackett  Laboratory, \\
Imperial College,\\ London SW7 2BZ, U.K.}\\
\vskip 0.8cm
and\\

\vspace*{7mm} {The Institute for Mathematical Sciences}\\
{\em Imperial College London} \\
{\em 53 Prince's Gate, London SW7 2PE, U.K.} \\

\end{center}
\vskip 1.0cm

\begin{abstract}\noindent

The doubled formulation of string theory, which is T-duality covariant  and enlarges spacetime with extra coordinates conjugate to winding number, is reformulated and its geometric and topological features  examined. It is used to formulate string theory in T-fold backgrounds with T-duality transition functions and  a quantum implementation  of the constraints of the doubled formalism is presented. This establishes the quantum equivalence to the usual sigma-model formalism for world-sheets of arbitrary genus, provided a topological term is added to the action. The quantisation  involves  a local choice of polarisation, but the results are independent of this. The natural  dilaton   of the doubled formalism is  duality-invariant  and so T-duality is a perturbative symmetry for the perturbation theory in the  corresponding   coupling constant. It is shown  how this  dilaton    is related to the dilaton  of the conventional sigma-model which does transform under T-duality. The generalisation of the doubled formalism to the superstring is  given and  shown to be equivalent to the usual formulation. Finally, the formalism is   generalised to one in which the whole spacetime is    doubled.

\end{abstract}

\vspace{1cm}



\vfill

\setcounter{footnote}{0}
\def\thefootnote{\arabic{footnote}}
\newpage

\section{Introduction}\label{Introduction}

A conventional \lq geometric' string background consists of a spacetime manifold equipped with a metric 
and various gauge fields, which may be connections for bundles or gerbes over spacetime, and satisfying field equations arising from the requirement that quantising  the corresponding sigma-model gives  a conformal field theory.
However, string theory can be consistently defined in many non-geometric backgrounds
that are not of this type \cite{Hull:2004in}-\cite{Lawrence:2006ma}, and it seems likely that generic string theory solutions
will be non-geometric.
In particular, 
conventional compactifications can be generalised to ones where the
internal compact manifold is replaced with string theory in a non-geometric background, 
resulting in a conventional theory in a geometric  four dimensional spacetime.
This has been explored in \cite{Dabholkar:2005ve}, where it was argued that this gives a much wider class of effective four-dimensional field theories than can be obtained from conventional compactifications.

An important class of non-geometric backgrounds are those which are  constructed from local
  patches, each of which   is a patch of a conventional geometric string background, but 
  these patches are glued together with transition functions that include duality transformations as well as the usual diffeomorphisms and gauge transformations \cite{Hull:2004in}.
 This can give T-folds with  T-duality transition functions or U-folds  with U-duality transition functions,
  or  mirror-folds  with mirror symmetry transition functions.
T-folds or U-folds require each patch to be the product of a torus with some open set in a base space $N$, so that the T-fold has a torus fibration over $N$, while a mirror-folds have a Calabi-Yau fibration.
  More exotic possibilities   include gluing 
 a heteroic string theory patch with a $T^4$ fibration to a $IIA$ string theory patch with a $K3$ fibration, as these theories are dual \cite{Hull:1994ys}.

The T-fold backgrounds can be studied within perturbative string theory and so can be most fully treated.
Locally, a T-fold looks like a conventional patch of a  spacetime with a torus fibration.
T-duality \cite{Giveon:1994fu} was shown in \cite{buscher},\cite{Rocek:1991ps},\cite{Giveon:1991jj},\cite{Alvarez:1993qi}
to be a symmetry of spacetimes that  torus fibrations in which there was a $U(1)^d$
isometry, so that they are principle $U(1)^d$ bundles. This was generalised in \cite{Hull:2006qs}
 to the case of
 general torus bundles in which there may be  no globally defined killing vectors, so establishing the   result  that T-duality  can be done fibrewise, provided that certain obstructions are absent. However, applying T-duality to   geometric backgrounds 
 with fluxes in general gives a T-fold \cite{Hull:2004in}, not a geometric space, and so one is led to consider such backgrounds.

Let $X^i$ be coordinates on the torus fibres, and $Y^m$ be the remaining coordinates, and   the $d^2$ moduli  $\t \in O(d,d)/O(d)\times O(d)$ of the torus $T^d$ depend  on $Y$ in general.
Quantising the 
coordinates $X$ gives a torus conformal field theory specified by the moduli $\t(Y)$ for each $Y$.
The conformal field theory has an $O(d,d;\Z)$ symmetry, and 
 moduli related by an $O(d,d;\Z)$ transformation determine the same conformal field theory.
  Then $O(d,d;\Z)$ transition functions  allow the consistent construction of a bundle of torus conformal field theories over some base space $N$ with local coordinates $Y$.
  One can then integrate over the fields $Y^m$ to give the quantum 
   string theory in such a T-fold background.
  
  In formulating the conformal field theory on the $T^d$ fibres, an extra $d$ coordinates $\ti X_i$
  for a dual torus $\ti T^d$ are needed.  These are conjugate to the winding number, and are needed to write vertex operators
  such as $e^{ik_L \cdot X_L}$ where $X_L=X-\ti X$, and 
 to formulate string field theory.
For string field theory in    toroidal backgrounds, the string field should depend explicitly on $\ti X$ as well as $X$ \cite{Kugo:1992md}. This means that generic  solutions of string field theory   depend on both $X$ and $\ti X$; some interesting examples of   backgrounds depending non-trivially on $\ti X$ have been investigated in \cite{Dabholkar:2005ve}.
However, T-fold backgrounds do not depend 
explicitly on $\ti X$, so can be expressed in terms of conventional spacetime fields locally. 
In \cite{Hull:2004in}, a formulation of string theory on a T-fold was given, with 
a target space which had a $T^{2d}$ doubled  torus fibration
with local coordinates $X^i, \ti X_i,Y^m$.
For a T-fold, the doubled patches fit together to form a 
$T^{2d}$ bundle $\hat M$  over the base
$N$, and the theory is formulated as a sigma-model with target space  $\hat M$.
This formulation is manifestly $O(d,d;\Z)$ invariant.
To obtain the conventional theory, a constraint is imposed that
halves the doubled degrees of freedom on the torus; for a flat background, this constraint requires half of the $2d$ scalar fields on $T^{2d}$ to be left-movers and half right-movers.
The constraint is well-defined on $\hat M$ and is   $O(d,d;\Z)$ invariant.
The conventional theory is regained by choosing a polarisation, i.e. by choosing half of the coordinates
on the torus $T^{2d}$ to be the physical spacetime coordinates.
This involves choosing a $T^d\subset T^{2d}$ and can be done globally for 
a geometric background, but only locally in each patch for a T-fold, and in general the 
polarisation changes from patch to patch.
T-duality can be thought of as acting to change the polarisation \cite{Hull:2004in}, and so the statement that the physics is T-duality invariant implies that the choice of polarisation does not affect the physics.

The sigma-model on the doubled space $\hat M$ can be quantised in the usual way, but the problem arises as to how to implement the constraint.
One approach is to first quantise the variables $X,\ti X$, for fixed $Y$.
One can first solve the constraint and then quantise.
The constraint  is a self-duality condition that relates $\pa \ti  X $ and $\pa X$, and it is important that
in the doubled formulation for a T-fold, $\ti X$  only enters through 
its derivative $\pa \ti  X $.
Then the constraint  can be used to give $\pa \ti  X $ in terms of $\pa X$.
The constraint implies the  classical world-sheet  field equations for $X,\ti X$, and 
for a cylindrical world-sheet
the field equation for $X$ can be solved in terms of   the oscillators, momenta
and winding modes for $X$. These can be quantised in the usual way to obtain the usual 
CFT on $T^d$. This gives a torus CFT with moduli $\t(Y)$ for each point $Y$ and hence a bundle of CFT's over $N$. The final step is then to quantise $Y$.

While this paper was in preparation, the paper  \cite{Hackett-Jones} appeared, giving  a constrained Hamiltonian approach   for T-folds on cylindrical world-sheets, using Dirac brackets to quantise the system.
This was   applied to an example of a  T-fold which is an asymmetric orbifold, and gave the same results as the conventional quantisation of this system. An interesting feature is that, at least for this explicit example, no choice of polarisation is needed.

However, it is desirable to have an off-shell formulation which does not impose field equations, and which applies to world-sheets that are Riemann surfaces of arbitrary genus.
The constraint requires that a certain conserved current $J$  vanishes, and it was suggested in \cite{Hull:2004in} that this could be imposed by gauging the symmetry generated by $J$, adding a coupling $C\cdot J$ to a world-sheet gauge field, plus quadratic terms in $C$. It will be shown here that this does not quite work, but that one can instead gauge half of the currents $J$ and this is sufficient to impose the constraint.
Gauge-fixing and integrating out the gauge fields then recovers the usual (undoubled) sigma-model formulation locally.
The choice of which  half of the currents $J$  to gauge is the choice of polarisation.
For a geometric background, there is a global choice of polarisation 
and the usual formulation is recovered, but for a 
T-fold, there is no global choice,
and the quantisation involves a choice of a different polarisation in each patch.
Nonetheless, the resulting quantum theories should patch together to give a consistent well-defined theory.
Then the situation is similar to gauge theory, which has a globally-well defined gauge-invariant quantum effective action, even though in the calculation of this one must  gauge-fix, breaking the manifest symmetry, and 
in general one must make a different gauge choice in different patches.

This allows the definition of the quantum theory for Riemann surfaces of arbitrary genus, and it is found that the classical action must be supplemented by a topological term in order to achieve complete equivalence to the usual formulation. This term does not affect the classical theory, but introduces certain relative signs in the sum over topological sectors.
It is also shown that there is a functional Jacobian  that arises in changing between the formulations, and this has important physical consequences at one-loop and higher.

For a T-fold to be a good string background, the resulting quantum theory must be conformal and modular invariant.
Conformal invariance requires that in any patch, $g,b$ and the dilaton must satisfy the usual $\b$-function equations, so that there is a conformal field theory in each patch. Modular invariance then imposes   conditions on the allowed  transition functions.
For example, a special class of T-folds are asymmetric orbifolds, and it is well-known that modular invariance
only allows a restricted class of asymmetric orbifolds.
Then a T-fold string background is locally conformal, i.e. it is constructed from patches in each of which 
the geometric data satisfies $\b$-function equations, and  the transition functions are chosen to be compatible with modular invariance.

In addition to showing how to quantise in the doubled formalism and establishing its equivalence to the usual formalism, a number of other issues left over from \cite{Hull:2004in} will be discussed.
The doubled formulation will be re-expressed in a form in which its geometric structure is more apparent, using results from  \cite{Hull:2006qs}.
A careful treatment of the global structure will be given and applied to the quantum theory.
A puzzle arises in  the issue of the dilaton coupling.
In the doubled formalism, the natural dilaton coupling through a Fradkin-Tseytlin term is necessarily duality invariant, while it is known that the dilaton in the 
usual sigma-model transforms under duality.
It will be shown that these results are consistent and that the dilatons in the two formalisms are indeed different, and the relationship between them will be found.
The string  perturbation theory involving the dilaton arising in the doubled formalism is duality invariant so that T-duality is manifestly a perturbative symmetry, and this coupling constant is the same as that of string field theory \cite{Kugo:1992md}.

In section \ref{Quantisation}, the  results are extended to the supersymmetric doubled formalism, and the relationship to the usual formalism again established.
In section \ref{Everything}, the formalism is   generalised to one in which all coordinates are doubled, not just the tori, and this gives a formalism applicable to general spaces, not just to torus bundles.

There is an interesting relation with Hitchin's generalised geometry \cite{Hitchin:2004ut}.
In generalised geometry, a conventional geometry with a $D$-dimensional  manifold $M$ equipped with a metric tensors $g$ and a gerbe connection $b$ is considered, and it is found that many features are elegantly expressed on
$T\oplus T^* (M)$, or  the twisting of this by a gerbe, and there is a natural action of the continuous group $O(d,d)$. The transition functions are 
diffeomorphisms and $b$-field gauge transformations, giving transition functions
$GL(D,\R)$ on $T\oplus T^* (M)$, or the semi-direct product of this with $b$-transformations for the twisted version.
T-folds are more general than generalised geometry, with transition functions including the discrete group 
$O(d,d;\Z)$ for $T^d$ fibrations, and are not manifolds with tensor fields 
$g,H$.
While generalised geometry doubles the tangent space, 
doubled geometry doubles the torus fibres, or the whole manifold. Doubling the manifold of course  entails doubling the tangent space. Both kinds of geometry have a natural action of $O(d,d)$ and similar $O(d,d)$ covariant structures appear in both.
However, doubled geometry is governed by the discrete group $O(d,d;\Z)$ and T-duality is an essential feature, while
in generalised geometry only the continuous group 
$O(D,D)$ appears.
On the other hand, the generalised geometry approach can be applied to any manifold, while
T-folds arise naturally only for torus fibrations.
The relation between doubled geometry and generalised geometry  will be discussed further elsewhere.

\section{String Backgrounds}\label{Back}

The string backgrounds that will be considered here  can be constructed from local patches and in each patch there is a conventional string background, so that each patch is diffeomorphic to a contractible open set in
$\R^D$ equipped with a metric $g$ and a 2-form $b$.
 A geometric background is a manifold made from patches of this type with transition functions that are diffeomorphisms and 2-form gauge transformations $\d b= d\l$, so that
 $g$ and $H=db$ are tensor fields on $M$.
 T-folds are non-geometric backgrounds where the 
 transition functions also include T-dualities, so that the result is not a manifold with tensor fields.
In this section, the local structure of such backgrounds will be reviewed, and the global structure will be discussed in section \ref{T-Folds}.

A geometric string background is then a manifold $M$ with  a metric $g$ and closed 3-form $H$.
In each local patch, one can introduce local coordinates
$\f ^\m$    ($\m,\n=1,..,D$, where $D$ is the dimension of $M$) and $H$ is given in terms of a 2-form potential $b$,
$H=db$. The lagrangian is
\begin{equation}
\label{lungage}
L=\ha \gmn d \f^\m \we  * d\f^\n + \ha \bmn  d \f^\m \we   d\f^\n\end{equation}
Here $d \f $ is a 1-form on $M$ pulled-back to the world-sheet. The
  world-sheet metric is taken to have   Lorentizan signature, and 
$*$  is the world-sheet  Hodge duality operator satisfying $(*)^2=1$.
(The formulae will be presented here for Lorentizan  world-sheet  metrics. The continuation to 
Euclidean  signature  is straightforward, and in most formulae 
in this paper is given by  
replacing $*$ with $-i*$, as $(-i*)^2=1$ in Euclidean  signature, and taking lagrangian 2-forms   
$L \to -i L$. In (\ref{lungage}), this has the net effect of replacing $b$ with $i\, b$.)

If $M$ is  a torus bundle   over some base manifold $N$
with fibres $T^d$, then
it can be constructed from 
 patches of the 
form $U'=U\times T^d$ where  $U$ is a patch on the base manifold $N$,
 diffeomorphic to a contractible open set in
$\R^{D-d}$.
In each such patch $U'$, there are
$d$ commuting vector fields
$k_i=k_i^\m \pa /\pa\f ^\m$ 
tangent to the fibres,  with
each $k_i$ generating a periodic orbit. 
It will be assumed they are Killing vectors
  with
\begin{equation} \label{LH}
{\cal L} _i H=0
\end{equation}
where $\cl _i$  is the Lie derivative with respect to $k_i$, generating a freely acting
$U(1)^d$ isometry of $U'$.
For principle bundles, these extend to globally defined Killing vector fields on $M$, but for general torus bundles they do not.
In \cite{Hull:2006qs}, T-duality and the gauging of sigma-models was generalised to such general torus bundles without isometries.

Consider then a patch of a string background $U'=U\times T^d$
with a metric $g$, a 2-form $b$ and $d$ Killing vectors in $U'$ tangent to the fibres.
They could fit together  to form either a torus bundle over $N$, or a T-fold over $N$.
The norm of the Killing vectors
\begin{equation}
\label{}
G_{ij} = g(k_i, k_j)
\end{equation}
defines a matrix of functions on $U$
 and, as this is non-degenerate in $U$ (assuming $g$ restricted to the fibres is positive definite), there
 are one-forms $\x^i$ with components
\begin{equation} \label{xis}
\x^i _\m=G^{ij} g_{\m\n} k_j^\n
\end{equation}
dual to the Killing vectors.
The field strengths
\begin{equation} \label{abc}
F^i=d\x ^i
\end{equation}
satisfy
\begin{equation} \label{abc}
\i_i  F^j=0
\end{equation}
where $\i_i$ deontes contraction with $k_i$.
The metric can be written as
\begin{equation} \label{gis}
g = \bar g +   G_{ij}\, \x^i \otimes \x^j
\end{equation}
where  $\bar g$ is 
  a  metric on $U $. The $\x ^i$ define a natural frame on the fibres over $U$.

Next we give an alternative derivation the results of \cite{Hull:2006qs} for the general form of $H$.
The condition (\ref{LH}) implies that $\i_ {j_1}...\i_{j_n}H$ is closed for $n=1,2,3$, so that
in a contractible open set $V\subset M$
they are exact. 
Then 
\begin{equation}
\label{}
\i_i \i_j \i_k H=K_{ijk}  
\end{equation}
are constants (in $V$)
and
\begin{equation}
\label{}
K=\frac 16 K_{ijk}  dX^i\we dX^j\we dX^k
\end{equation}
defines a closed 3-form, so that
\begin{equation}
\label{}
H'=H-K
\end{equation}
is closed and satisfies
\begin{equation}
\label{iiih}
\i_i \i_j \i_k H'=0  
\end{equation}
The analysis of \cite{Hull:2006qs} can now be applied to $H'$.
In $V$, there is are 1-forms $v_i$ and 0-forms $B_{ij}=- B_{ji}$
such that
\begin{eqnarray}
\i_i \i_j  H'&=&-dB_{ij} \\
\i_i  H'&=& dv_i \end{eqnarray}
and (\ref{iiih})
implies
\begin{equation}
\label{}
\cl_i B_{jk}=0
\end{equation}
The 1-forms $v_i$ are only defined up to the addition of an exact 1-form.
Consider then
\begin{equation}
\label{}
v'_i= v_i- d f_i
\end{equation}
where $f_i$ are functions on $V$ satisfying
\begin{equation}
\label{sdfgg}
\i_i d f_j = B_{ij}+ \i_i v_j
\end{equation}
The integrability condition 
$\i_k d \i_i d f_j =\i_id\i_k d f_j$
for (\ref{sdfgg}) is satisfied as a result of (\ref{iiih}) \cite{Hull:2006qs}, so that solutions $f_i$ exist.
Then
\begin{equation}
\label{ivp}
\i_i v'_j= B_{ij}
\end{equation}
and 
\begin{equation}
\label{clvp}
\cl _i v'_j=0
\end{equation}
 The locally-defined 1-forms
 \begin{equation}
\label{}
\ti A_i=v'_i+B_{ij}\x ^j
\end{equation}
are horizontal
\begin{equation}
\label{}
\i_i \ti A_j=0
\end{equation}
and invariant
\begin{equation}
\label{}
\cl _i \ti A_j=0
\end{equation}
so that they 
can be regarded as 1-forms on   $U\subset N$.
They are connections for a bundle over $N$ \cite{Hull:2006qs}  with curvature
\begin{equation}\label{ftiiss}
\ti F_i = d \ti A_i
\end{equation}
which is horizontal, $\i_i \ti F_j=0$.
Then 
\begin{equation} \label{hiso}
H=\bar H +  \ti F_i \we   \x^i + dB +K
\end{equation}
where
\begin{equation}\label{biso}
B=\ha   B_{ij}
   \x ^i \we  \x ^j
\end{equation}
and
$\bar H$
  is a 3-form on $N$ satisfying
 \begin{equation}
d\bar H= - \ti F_i \we F^i;
\end{equation}
A 2-form potential $b$ with $db=H$ is given by
   \begin{equation}
\label{bisoo}
b= \bar b +  \x^i \we  \ti A_i+ \ha   B_{ij}   \x ^i \we  \x ^j
 +\k
\end{equation}
where $d \k=K$, so that $\k$ can be taken to be
\begin{equation}
\label{}
\k= \frac 16 K_{ijk}  X^i\we dX^j\we dX^k
\end{equation}
$\bar b $ is a 2-form on $U\subset N$ with
\begin{equation}
\label{}
\bar H= d \bar b + F^i\we \ti A_i
\end{equation}
Using the symmetry, these results extend from a contractible patch to any patch of the form $U'=U\times T^d$.

In adapted local coordinates 
$\f^\m= (X^i, Y^m)$ in which 
\begin{equation} \label{abc}
k_i^\m \frac {\pa} {\pa \f^\m}= \frac {\pa} {\pa X^i}
\end{equation}
the Lie derivative is the partial derivative with respect to $X^i$, so that
$\gmn,\hmnp$ are independent of $X^m$.
Then 
\begin{equation} \label{aiss}
\x^i = dX^i + A^i
\end{equation}
where
$A^i=A^i_m (Y)d Y^m$ satisfies
$\i _i A^j=0 $ and
\begin{equation} \label{abc}
dA^i=F^i
\end{equation}
satisfies 
$\i_i F^j=0$.
The $A^i $ are connection 1-forms for $M$ viewed as a bundle over $N$.

The derivation of T-duality of \cite{Hull:2006qs}, generalising   that of \cite{buscher},\cite{Rocek:1991ps},
\cite{Giveon:1991jj},
\cite{Alvarez:1993qi},
involves the gauging of the symmetry generated by the $k_i$.
If $K=0$ and the $\ti A$    and the $B_{ij} $ 
  are globally defined,  
  then
the obstructions to gauging of \cite{Hull:1989jk} are absent as a result of (\ref{ivp}),(\ref{clvp}); however, in general $\ti A$    and   $B_{ij} $ will not be globally defined.
  
  In general the $\ti A_i= \ti A_{im }(Y)dY ^m $ are connections for a dual bundle  $\ti M$ over $N$, built from
 patches $U\times \ti T^d$,
  and so will not be globally defined.
Globally defined one-forms are defined by
introducing fibre coordinates $\ti X_i$ on $ \ti T^d$ 
so that
\begin{equation}
\ti \x_i = d\ti X_i+ \ti A_i 
\end{equation}
is a well-defined 1-form on $\ti M$
and
\begin{equation}
\ti F_i = d\ti \x_i = d\ti A_i
\end{equation}
is also horizontal. To be able to define a well-defined quantum sigma-model, the fibres   $\ti  T^d$ are taken to be 
the torus   dual to the torus fibres in $U'=U\times T^d$ \cite{Hull:2006qs}. 
If $X^i$ has period $2\p R ^i$ and  $\ti X^i$ has period $2\p \ti R _i$, then these are related by
$R^i=\a'/ \ti R_i$.
In addition to the $k_i, $ there
are
 vector fields $\ti k^i$ 
tangent to the new fibres
\begin{equation}
\label{}
\ti k^i = \frac {\pa }{\pa \ti X_i}
\end{equation}
The $\ti k^i$ commute with the $k_i$.

This allows the construction of a doubled 
patch $\hat U= U\times T^d\times \ti T^d$ with
fibres $T^{2d}$,  coordinates $Y^m$ and 
\begin{equation}
\label{xcol}
 \cx ^I=
\left( \begin{array}{c} X^i\\ \ti X_i \end{array}\right)
 \end{equation}
where $I=1,...,2d$, and connections
\begin{equation}
\label{ }
 \ca ^I =
\left( \begin{array}{c} A^i\\ \ti A_i \end{array}\right)
\end{equation}
so that
the one-forms
\begin{equation}
  \Xi =
\left( \begin{array}{c} \x^m\\ \ti \x_m  \end{array}\right)
\label{xicol}  \end{equation}
are well-defined 1-forms.

The isometries on $\hat M$   can now be gauged  provided   there is no 3-flux on the fibres \cite{Hull:2006qs}:
\begin{equation} \label{iiihis}
\i_i \i_j \i_k H= 0
\end{equation}
so that $H'=H$, and this will be assumed to be the case here.
This is also the condition for
 conventional T-duality to be possible  \cite{Hull:2006qs}.

For  geometric  backgrounds, the patches $U'= U\times T^d$ patch together to give a 
manifold $M$, the dual patches $\ti U= U\times \ti T^d$
patch together to form a dual manifold $\ti M$ (the T-dual of $M$, again a torus bundle over $N$) and the 
$\hat U= U\times T^d\times \ti T^d$  patch together to form a manifold $\hat M$, which is
a $T^{2d}$ bundle over $N$.
For T-folds, the $U'$ or $\ti U$ may not patch to form manifolds, but  $\hat M$
is a well-defined $T^{2d}$ bundle over $N$, a geometric space containing all the information about the background and all its T-duals.
It is this well-defined manifold $\hat M$ that is used to construct the string action for a T-fold background using the doubled formalism \cite{Hull:2004in}.

There is a natural action of $O(d,d)$ on $\hat U$ and hence on $\hat M$.
Consider   $h\in O(d,d)$ given
by
 \begin{equation}\label{oddm}
h=\left(
\begin{array}{cc}
a & b\\
c & d \end{array}\right),
  \end{equation}
where $a,b,c,d$ are $d\times d$ matrices.
This preserves the indefinite metric  \begin{equation}
L=\left(\begin{array}{cc}
0 & \1 \\
\1 & 0
\end{array}\right)
\label{liss}  \end{equation}
 so that
\begin{equation}
h^tLh=L\;\; \Rightarrow \;\;a^t c+c^t a=0,\;\;\;b^t d+d^t
b=0,\;\;\;\ a^t d+c^t b=\1 . \label{abc}  \end{equation} 
The group $O(d,d,\Z)$ consists of matrices (\ref{oddm}) with integral entries.
Then $\Xi, \cx, \ca$ transform in the fundamental representation
\begin{equation}
\label{xxitrans}
\Xi \to \Xi ' = h^{-1} \Xi
\end{equation}
\begin{equation}
\label{atranss}
\ca \to \ca' = h^{-1} \ca, \qquad \cx \to \cx' = h^{-1} \cx
\end{equation}

Defining 
\begin{equation}
\label{ }
E_{ij}=G_{ij}+B_{ij}
\end{equation}
$E$ transforms non-linearly under $O(d,d)$
 \cite{GMR},\cite{Giveon:1991jj},\cite{Giveon:1994fu},\cite{Hull:2006qs}
\begin{equation}
E'  = (aE+b)(cE+d)^{-1}. \label{tetrans}  \end{equation} 
The moduli $G,B$ can be used to define a natural metric on the fibres given by the
$2d\times 2d$ matrix $\cm _{IJ}$ given by
\begin{equation}
\cm =\left(
\begin{array}{cc}
G-BG^{-1}B & BG^{-1} \\
-G^{-1}B   & G^{-1}
\end{array}
\right).
\label{genmet}  \end{equation}
which transforms covariantly under $O(d,d)$
\begin{equation}
\cm \to h^t \cm h
\label{gtrans}  \end{equation}
Note that the $G,B$ are well-defined moduli and are scalar fields on $N$, so that the
metric (\ref{genmet})
and the transformations (\ref{tetrans})
are well-defined.
Similar formulae involving the components of the gauge field $b$ are potentially  problematic as $b$ is only defined up to gauge transformations.

\section{Doubled Formalism}\label{doubled}

The doubled formalism \cite{Hull:2004in} 
is based on the duality-covariant  formalism of \cite{Cremmer:1997ct}
 (and similar to   models of \cite{Kugo:1992md},\cite{Tseytlin:1990nb},\cite{Tseytlin:1990va},\cite{Maharana:1992my},\cite{Hull:1988dp},\cite{Hull:1988si},\cite{Duff:1990}).
It is $O(d,d;\Z)$ covariant and written in a patch $\hat U$ of $\hat M$  in terms 
of $\cx, \ca ,\cm$. The usual formalism arises on choosing a polarisation, i.e. a choosing 
 a physical   subspace $U\times T^d \subset U\times T^{2d}$.

Consider a patch $\hat U$ of
a space $\hat M$ which is a $T^{2d} $ bundle over $N$, with
fibre coordinates $\cx^I$, local coordinates $Y^m$  
on    $U\subset N$ and connection 1-forms
     \begin{equation}
\ca^I=\ca^I_mdY^m 
\end{equation}
 Let $L_{IJ}$ be the constant $O(d,d)$ invariant metric (\ref{liss}) on the fibres, and 
 let $\cm_{IJ}$ be a positive-definite fibre metric satisfying
 \begin{equation}
\label{hcons}
L^{-1}\cm L^{-1}\cm =\1
\end{equation}
This \lq generalised metric' is assumed to be indepedent of $\cx$  but is a  function  $\cm _{IJ}(Y)$  on $N$.
Then
 \begin{equation}
S^{I}{}_J= L^{IK}\cm _{KJ}
\label{abc}  \end{equation}
satisfies 
 \begin{equation}
S ^2 = \1
\label{abc}  \end{equation}
and so defines an almost product (or almost real) structure.

The sigma-model with target space $\hat M$ of \cite{Hull:2004in}   is a theory of maps from a 2-dimensional world-sheet $W$ to $\hat M$, given locally by $\cx^I(\s)$ where $\s^\a$ are coordinates on $W$.
The pull-back of $d\cx^I$ gives the fibre momentum
\begin{equation}
  \cp ^I _\a =\pa _\a \cx ^I
\label{abc}  \end{equation} 
while 
the pull-back of the one-forms $\Xi^I$ gives  the covariant fibre momentum $\hat \cp ^I$, which is a  1-form on 
$W$ with components
\begin{equation}
\label{}
\hat \cp_\a ^I = \cp _\a ^I+\ca ^I _m \pa _\a Y^m
\end{equation}
The lagrangian   of \cite{Hull:2004in} is
\begin{equation}
\cl _d= \frac 1 4
 \cm_{IJ} \, \hat  \cp ^I \we *\hat  \cp ^J - \half L_{IJ} \cp ^I\we \ca ^J  +\cl (Y) \label{lagdh}  \end{equation}
where $\cl (Y)$ is the lagrangian for a sigma-model with target space $N$, and all forms have been pulled back to $W$.
The unusual normalisation with a factor of $1/4$ is important and needed to 
give equivalence with the canonically normalised
 standard sigma-model lagrangian (\ref{lungage}).
 The Wess-Zumino term
\begin{equation}
\label{}
 S_{WZ} =- \ha \int _W L_{IJ} \cp ^I\we \ca ^J  
\end{equation}
  can be rewritten as
\begin{equation}
\label{}
 S_{WZ} =- \ha \int _V L_{IJ} \cp ^I\we \cf ^J  
\end{equation}
where $V$ is a 3-manifold with boundary $W$, and $ \cf ^I$ is the pull-back  $$  \ha \cf ^I_{mn} \pa _\a Y^m   \pa _\b Y^n d\s^\a \we d \s ^\b$$ of the curvature
\begin{equation}
\label{}
\cf^I=d\ca ^I
\end{equation}
In later sections, it will be useful to consider adding a toplogical term
\begin{equation}
\label{topo}
\cl_{top}= \ha \W _{IJ} d\cx^I\we d\cx^J
\end{equation}
for some constant $\W _{IJ}=- \W _{JI}$.
This does not contribute to the field equations and does not affect the classical theory, but plays a role in  the quantum theory.

This theory is subjected to the constraint \cite{Hull:2004in}
\begin{equation}
\label{scons}
\hat \cp = S*\hat \cp  
\end{equation}
where
$*$ is the Hodge dual on the world-sheet
 satisfying $(*)^2=1$  (assuming Lorentzian signature world-sheet; for $W$ with  Euclidean signature,
 the constraint is $\hat \cp = -iS*\hat \cp  $.)
If the sigma-model on  $N$
has a lagrangian
\begin{equation}
\label{lagyp}
\cl (Y)
=\cl '(Y)
-   A^i\we A_i
\end{equation}
where
  \begin{equation}
\label{lagy}
\cl ' (Y)= \ha \bar g_{mn} dY^m \we * dY^n+\ha \bar b_{mn} dY^m \we  dY^n
\end{equation}
for some $ \bar g_{mn}(Y),  \bar b_{mn}(Y)$ on the base $N$,
then it was shown in \cite{Hull:2004in}
that the doubled sigma-model (\ref{lagdh}) with constraint (\ref{scons})   is classically equivalent
to the conventional  sigma-model (\ref{lungage}) with metric (\ref{gis}) and 2-form (\ref{bisoo}).
 In section \ref{Constraint}, this result will be re-derived and extended to the quantum theory in section \ref{Quantisation}.

The field equation from varying $\cx ^I$ in (\ref{lagdh}) is
\begin{equation}
\label{field}
d*(\cm _{IJ} \hat P^J)= L_{IJ} \cf ^J
\end{equation}
which can be rewritten as 
\begin{equation}
\label{}
  d*(S^I{}_J \hat P^J- *\hat P^I)=0
\end{equation}
so that  the constraint  (\ref{scons}) implies the field equation (\ref{field}) (and is a stronger condition).
 
The lagrangian is manifestly invariant under the  rigid  $GL(2d, \R)$
transformations
\begin{equation}
\cm \to h^t \cm h, \qquad \cp \to h^{-1} \cp, \qquad \ca \to h^{-1} \ca
\label{gtrans}  \end{equation}
(with $Y$ and $\cl (Y)$ invariant).
The corresponding transformation of the coordinates
\begin{equation}
  \cx \to h^{-1} \cx
\label{xtrans}  \end{equation}
only preserves the boundary conditions
if $g$ is restricted to be in the subgroup $GL(2d,\Z) \subset GL(2d, \R)
$ preserving the periodicities of the $\cx$.
The constraint (\ref{scons}) breaks $GL(2d, \R)$
   to the subgroup $O(d,d)$ preserving $L^{IJ}$ and so breaks $GL(2d,\Z)$ to
   $O(d,d;\Z)$.
   Thus this formulation is manifestly invariant under the T-duality group
      $O(d,d;\Z)$.
      The topological term (\ref{topo}) is invariant if 
      $\W \to h^t \W h$ under these transformations.

\section{Polarisation and T-Duality}\label{polar}

In
order to make contact with the conventional formulation, one needs
to choose a polarisation, i.e. to choose a splitting of $T^{2d}$
into a physical $T^d$ and a dual $\ti T ^d$ for each point in $N$,
splitting the fibre coordinates into the physical coordinates
$X\in T^d$ and the dual coordinates $\ti X\in \ti T^d$, and then
write the theory in terms of the coordinates $X$ alone, solving
the constraint (\ref{scons}) to express $\ti X(\s)$ in terms of $X(\s)$.
Then the variables $X$ are   the ones integrated over in the
functional integral, and invariance of the theory under T-duality
implies that the  physics should be independent of the choice of
polarisation.

In order to define a polarisation or local product structure on
the fibres,  one first chooses a subgroup $GL(d,\R)$ of $O(d,d)$ under
which the fundamental ${\bf 2d}$ of $O(d,d)$ splits into  the
fundamental representation ${\bf d}$ of $GL(d,\R)$  and the dual
representation ${\bf d'}$, ${\bf 2d} \to {\bf d} \oplus {\bf d'}$. It will be useful to
use a superscript $i$ for the fundamental representation ${\bf d}$
(where $i=1,..., d$) and a subscript $i$ for the  dual
representation ${\bf d'}$, and 
introduce constant projectors $\P ^i
{}_I$   and    $\Tilde \P _{iI} $, so that
\begin{equation}
  \cp =\left( \begin{array}{c} \P ^i {}_I \cp ^I
    \\  \Tilde \P _{iI} \cp ^{I}  \end{array}\right)=
\left( \begin{array}{c} P  ^i  \\ Q _i  \end{array}\right).
\qquad \cx =\left( \begin{array}{c} \P ^i {}_I \cx ^I
    \\  \Tilde \P _{iI} \cx ^{I}  \end{array}\right)=
\left( \begin{array}{c} X  ^i  \\ \ti X _i  \end{array}\right),
\label{polar}  \end{equation} 
with the $X^i$ the coordinates of the
$T^d$ subspace and $\ti X_i$ the coordinates of the dual $\ti T^d$
subspace.
This can be thought of as a choice of
basis, but it is useful to introduce the projectors explicitly, so
as to keep track of the choice of subgroup  $GL(d,\R)$ of
$O(d,d)$; duality transformations change the
projectors and change the subgroup  $GL(d,\R)$ to a conjugate one.

The metric $L$ is off-diagonal in the $GL(d)$ basis  and can be written as
\begin{equation}
L=\left(\begin{array}{cc}
0 & \1 \\
\1 & 0
\end{array}\right)
\label{abc}  \end{equation}
so that the corresponding line element is
\begin{equation}
ds^2= 2 dX^i d\ti X_i \label{abc}  \end{equation} Then the
$T^d$ submanifold with coordinates $X^i$ is a maximally null
subspace with respect to this metric. Choosing a polarisation 
that selects a maximal null $T^d \subset T^{2d}$
together with its complement $\ti T^d$
  then corresponds to choosing a subgroup
$GL(d,\Z) \subset O(d,d;\Z)$.

It will be useful to introduce the notation
 $\hat I$ for the $O(d,d)$ indices in the $GL(d)$ basis, so that
for  any vector $v$, $v^{\hat I}=(v^i,v_i)$ and the matrix giving
the change from an arbitrary basis to the $GL(d)$ basis is
\begin{equation}
\qwe^{\hat I} {}_J= \begin{pmatrix} \P ^i{}_J \\  \ti \P  _{iJ}  \end{pmatrix}
\label{abc}  \end{equation}
 with the corresponding matrix for  the
dual representation
$\hat \qwe = L^{-1} \qwe L$ so that
\begin{equation}
\hat \qwe _{\hat I} {}^J= \begin{pmatrix} \ti \P  _{i}{}^J \\
\P ^{iJ }\end{pmatrix} \label{abc}  \end{equation} where $  \P
^{iJ }= \P ^i{}_I L^{IJ}$, $\ti \P  _{i}{}^J=\ti \P  _{iI}L^{IJ}$.
The matrix $\qwe^{\hat I} {}_J$ can be thought of as a
representative of the coset $O(d,d)/GL(d,\R)$, or as a \lq
vielbein' converting $O(d,d)$ indices to $GL(d)$ ones. 
Then the equations giving components in the $GL(d)$   basis can be
rewritten as
\begin{equation}
\qwe   \cp =
\left( \begin{array}{c} P  ^i  \\ Q _i  \end{array}\right), \qquad
\qwe   \ca   =\left( \begin{array}{c}A^i      \\ \ti A_i  \end{array}\right)
\label{jfiss}
  \end{equation}
The components of $\cm$ in this basis
\begin{equation}
\hat \qwe   \cm \hat \qwe   ^t=  \left(
\begin{array}{cc}
G-BG^{-1}B & BG^{-1} \\
-G^{-1}B   & G^{-1}
\end{array}
\right).
\label{hfis}
  \end{equation}
This notation will help in    following  the effects of changes of polarisation explicitly.
In particular, (\ref{hfis}) defines a metric $G_{ij}$ and 2-form $B_{ij}$ in terms of $\cm$ and
a polarisation $\qwe  $.

The T-duality transformation rules $G\to G'$, $B\to B'$, $\ca\to \ca '$
(\ref{tetrans}), (\ref{atranss}) are then obtained   using the
$O(d,d)$ transformations for $\cm, \ca$ while keeping the
polarisation $\qwe  $ fixed,
\begin{equation}
\cm \to \cm '= h^t \cm h,  \qquad \ca \to \ca ' = h^{-1} \ca,  \qquad
\qwe   \to \qwe  '= \qwe   \label{hact}  \end{equation}
 so that e.g.
\begin{eqnarray}
  G^{-1}&=& \P \cm  \P^{t} \quad  \to \quad  (G')^{-1}=\P  h^t \cm h  \P^{t}
  \nonumber \\
 B G^{-1}&=& \ti \P \cm  \P^{t}  \quad \to \quad B'(G')^{-1}= \ti \P  h^t \cm h  \P^{t}
 \nonumber \\
 A&=& \P \ca  \quad \to \quad A' =  \P  h^t \ca
 \end{eqnarray}
These same transformations $G\to G'$, $B\to B'$, $A\to A'$ can also be obtained by keeping $\cm$ fixed while transforming $\qwe  $
\begin{equation}
\cm \to \cm '=  \cm ,  \qquad \ca \to \ca ' =   \ca,  \qquad \qwe  
\to \qwe  '= \qwe   h \label{hpas}  \end{equation} so that
\begin{equation}
 \P  \to \P '= \P  h   \qquad  \ti \P  \to\ti  \P '= \ti  \P  h
\label{ppas}  \end{equation}

Thus the T-duality transformations can be viewed either as active
transformations in which the geometry $\cm, \ca$ is changed while
$\P,\ti \P$ are kept fixed (\ref{hact}), or as a passive one in which  the
geometry $\cm, \ca$ is kept fixed but the polarisation is changed
(\ref{hpas}),(\ref{ppas}). In the latter viewpoint, the doubled
geometry is unchanged, but the choice of physical subspace is
transformed. The symmetry under T-duality is then the statement
that the physics does not depend on the choice of physical
subspace.

\section{Conserved Currents}\label{Currents}

The one-forms on  the world-sheet $W$
\begin{equation}
\label{Jiss}
J_I=\cm _{IJ} \hat P^J- L_{IJ} *\hat P^J
\end{equation}
are conserved currents
\begin{equation}
\label{}
d*J_I=0
\end{equation}
(using the field equations (\ref{field})).
It is the sum of a Noether current $\cj_I$ and 
the \lq topological' current
\begin{equation}
\label{}
 j _I = L_{IJ} * \cp ^J
\end{equation}
which is trivially conserved, $d* j=0$ as $d\cp =0$.
The Noether current is
\begin{equation}
\label{}
\cj_I=\cm _{IJ} \hat P^J- L_{IJ} *\ca ^J
\end{equation}
where $ \ca ^J = \ca ^J_m  \pa _\a Y^m  d\s ^\a$ is the pull-back of $\ca$,
and this generates the  symmetries
\begin{equation}
\label{varx}
\d \cx^I=\a ^I
\end{equation}
of translation along the fibres.
Note that $J$ is  gauge-invariant and so well-defined, while $\cj $ and $j$ are not.
The constraint (\ref{scons}) is $J_I=0$. (Adding the topological term (\ref{topo})
would modify the Noether current by an identically conserved term,
$\cj _I \to \cj_I +\W_{IJ}  * \cp ^J$.)


Following \cite{Hull:2004in}, it is useful to introduce a $2d\times 2d$ vielbein $\cv ^A {}_I(Y)$  such that 
\begin{equation}
\cm = \cv ^t \cv \label{abc}  \end{equation}
with frame indices raised and lowered with $\d _{AB}$.
There are then two metrics, $\cm _{IJ}$ with frame components $\d _{AB}$ and 
$L_{IJ}$ with frame components $L_{AB}$. They are both preserved by $O(d)\times O(d)$, and it is useful to 
choose a basis in which  $O(d)\times O(d)$ is manifest.
The indices
$A,B=1,...,2d$ transform under $O(d)\times O(d)$ and can be split
into indices $a,b=1,...,d$ and $a',b'=1,...,d$ for the two $O(d)$
factors, $A=(a,a')$, so that in a natural basis
\begin{equation}
L^{AB} = \begin{pmatrix} L^{ab}& 0 \\ 0& L^{a'b'}
  \end{pmatrix} =
   \begin{pmatrix} \1 ^{ab}& 0 \\ 0& -\1 ^{a'b'}
  \end{pmatrix}, \qquad
  S^{A}{}_B = 
   \begin{pmatrix} \d ^a {}_b & 0 \\ 0& -\d ^{a'}{}_{b'}
  \end{pmatrix}
\label{abc}  \end{equation}
Then
\begin{equation}
\cv ^A {}_I =\left( \begin{array}{c} \cv  ^a {}_I \\ \cv ^{a'} {}_I \end{array}\right),
\qquad  \cv \cp =
\left( \begin{array}{c} \cp  ^a  \\ \cp ^{a'}  \end{array}\right),
\label{abc}  \end{equation}
and
\begin{equation}
\cm _{IJ}= \cv  ^a {}_I \cv  ^b {}_J \d _{ab} + \cv  ^{a'} {}_I \cv  ^{b'} {}_J \d _{a'b'}
\label{abc}  \end{equation}
The current 
\begin{equation}
\label{jcur}
J^I=L^{IJ}J_J=S^I{}_J \hat P^J- *\hat P^I
\end{equation}
has frame components  $J^A= (J^a, J^{a'})$
\begin{eqnarray}
J ^{a} &=&\hat \cp ^{a} - * \hat \cp^{a} 
\nonumber \\
J ^{a'} &=&\hat \cp ^{a'} +* \hat \cp^{a'} 
\end{eqnarray}
The constraint (\ref{scons}) is 
$J=0$ and this becomes
\begin{eqnarray}
\hat \cp ^{a} &=&+ * \hat \cp^{a} 
\nonumber \\
\hat \cp ^{a'} &=& -* \hat \cp^{a'} 
\end{eqnarray}
Introducing null coordinates $\s^\pm$  on the world-sheet, so that  $\a =(+,-)$, these become
 \begin{eqnarray}
\hat \cp ^{a}_- &=&0
\nonumber \\
\hat \cp ^{a'}_+ &=&0
\end{eqnarray}
while $J^a_\a, J^{a'}_\a$ are the chiral currents
\begin{eqnarray}
J^{a}_+ & =  0, \qquad J^{a}_- & =  \hat \cp ^{a}_-,   \nonumber \\
J^{a'}_+ & =  \hat \cp ^{a'}_+, \qquad J^{a'}_- & = 0 \label{jisss}
\end{eqnarray}
There are then two chiral currents, and their conservation law is
\begin{equation}
\label{}
D*J^A = d*J^A-\w^A{}_B \we * J^B=0
\end{equation}
where $\w $ is the connection $\w _\a= (\pa _\a \cv) \cv^{-1}$ and  has off-diagonal terms mixing the two currents.
For example, the conservation law for $J^{a}_-$ is
\begin{equation}
\label{}
\pa_+J^{a}_- - (\w _+)^a{}_b J^b_- - (\w _-)^a{}_{b'} J^{b'}_+=0
\end{equation}

Given a polarisation, one can define the currents
\begin{equation}
\label{Jissi}
J^i=  \P ^i{}_{I}J^I
\end{equation}
which are  conserved $d*J^i=0$ as  $  \P ^i{}_{I}$ is constant.
The components of $J^i_{\a}$
 are given, using (\ref{jisss}),  by
 \begin{equation}
\label{}
J^i_{+}  =     \P ^i{}_{a'}   \hat \cp ^{a'}_+
\qquad
J^i_{-}  =     \P ^i{}_{a}   \hat \cp ^{a}_-
\end{equation}
where
\begin{equation}
\label{fdghsh}
   \P ^i{}_{a} =  \P ^i{}_{I}\cv ^I_a, 
\qquad
   \P^i{} _{a'} = \P ^i{}_{I} \cv ^I_{a'}
\end{equation}
As the matrices (\ref{fdghsh})
  are non-degenerate,  $J^i=0$ is equivalent to $J^I=0$
 as the only non-vanishing components of $J^I$ are those in (\ref{jisss}), and so  $J^i=0$ is equivalent to  the constraint  (\ref{scons}).
 
As before, $J^i = \cj^i + j^i$ where  $j^i =  \P ^i{}_{I} * d\cx ^I$ is trivially conserved and $\cj^i$ is the Noether current for the transformations
\begin{equation}
\label{varxp}
\d \ti X_i=\ti \a _i , \qquad \d X^i=0
\end{equation}
Similarly, there are also conserved  currents 
\begin{equation}
\label{}
J_i= \Tilde \P _{iI}J^I
\end{equation}
  with $J_i=0$ equivalent to
(\ref{scons}) and which generate the transformations
$\d X^i=\a ^i, \d \ti X_i=0$.

In the case of a trivial bundle with constant $\cm_{IJ}=\d_{IJ}$, the currents are
\begin{equation}
\label{}
J^i = 
 d\ti X_i +*dX^i
 \end{equation}
(the flat metric can be used to identify upper and lower indices $i,j$ and tangent space indices $a, a'$).
The current
$d\ti X_i$ generates
$\d \ti X_i=\a _i$
while
$*dX $ is a topological current that is automatically conserved.
Similarly, the current
\begin{equation}
\label{}
J^i = 
dX^i +* d\ti X_i  
\end{equation}
is the sum of  a current $dX^i $ generating $\d X^i=\a^i$ and the topological current $*d\ti X$.
In the $O(d)\times O(d)$ basis
\begin{equation}
\label{xcola}
 \cx ^I=
\left( \begin{array}{c}X_R^a \\ X_L^{a'}  \end{array}\right)
 \end{equation}
with 
\begin{equation}
\label{}
X^i= \frac{1}{ 2} \left( X_L^i+X_R^i \right), \qquad
\ti X_i= \frac{1}{ 2} \left( X_R^i-X_L^i \right)
\end{equation}
Then 
\begin{equation}
\label{}
\cp ^a _\a = \pa _\a X^a _R, \qquad \cp ^{a'} _\a = \pa _\a X^{a'} _L
\end{equation}
and the currents (\ref{jisss}) are
\begin{eqnarray}
& J^{a}_+  =  0, \qquad &J^{a}_-  =  \pa _- X_R^a,     \\
& J^{a'}_+  = \pa _+
X_L^{a'}, \qquad & J^{a'}_-  = 0 
\end{eqnarray}
The   symmetries generated by $J^I$ 
are
\begin{equation}
\label{}
\d X_R^a= \a_R^a, \qquad \d X_L^{a'}= \a_L^{a'}
\end{equation}
and $J^i$ generates the anti-diagonal subgroup with 
$\a_L^i=-\a_R^i $ while $J_i$ generates the 
diagonal subgroup with 
$\a_L^i=\a_R^i $.
Note that
the currents generate a Kac-Moody algebra
\begin{equation}
\label{}
[J^a_-(\s), J^b_-(\s')]= d \d^{ab}\d ' (\s-\s'),
\qquad 
[J^{a'}_+(\s), J^{b'}_+(\s')]= d \d^{a'b'}\d ' (\s-\s'),
\end{equation}
and so $J=0$ is a second class constraint.
This means that it cannot be imposed by adding a lagrange multiplier term $C\cdot J$, but might  be imposed by 
supplementing this with  a further term involving $C^2$; this will be discussed in the next section.

The constraint (\ref{scons})  then implies $\pa _- X^a_R=0$ and $\pa _+
X^{a'}_L=0$ so that $X^a_R$ are right-movers and $X^{a'}_L$ are
left-movers, giving the right count of degrees of freedom. 
The generalisation of this to the interacting case is that the constraint  (\ref{scons}) implies that half of the
currents $J^I$ are chiral and the other half anti-chiral, but the 
projectors onto the  chiral and anti-chiral parts change with the coordinate $Y$, as they are given in terms of $S(Y)$.

\section{Imposing the Constraint}\label{Constraint}

The constraint (\ref{scons}) is $J^I=0$ where $J^I$ is the current (\ref{Jiss}).
Given a polarisation, the constraint $J^i=0$ where $J^i$ is the current (\ref{Jissi}) also implies (\ref{scons}).
A natural way of  imposing the constraint is to attempt to gauge the symmetries generated by the current $J^I$ or $J^i$, as suggested in \cite{Hull:2004in}. 
This involves introducing a gauge field $C_I$ or $C_i$ which is a one-form 
on the world-sheet.
The linear Noether coupling is then
\begin{equation}
\label{cj}
\frac 12  C_I\we *J^I
\end{equation}
or
\begin{equation}
\label{cji}
\frac 12  C_i\we *J^i
\end{equation}
so that if this were the only term involving $C$, the gauge field would be a lagrange multiplier 
imposing the constraint $J=0$.
However, gauge invariance requires adding a term quadratic in 
$C$.
Defining $C_A=(C_a, C_{a'})$ by $C_A=\cv_A{} ^I C_I$,
using (\ref{jisss}), the term (\ref{cj})
is
\begin{equation}
\label{}
\frac 12   \left(  C_+^a J_-^a + C_-^{a'}J_+^{a'} \right)
\end{equation}
and $C_-^a, C_+^{a'}$ do not appear, and as a result gives the same coupling as (\ref{cji}).
However, there are in addition terms quadratic in $C$; for the coupling to $J^I$, these 
     do depend on $C_-^a, C_+^{a'}$, while for the coupling to $J^i$, they do not.

The first step in the gauging of $J^I$ is given by minimal coupling, so that
$\cp^I$ is replaced with
\begin{equation}
\label{}
  \cp ^I +L^{IJ}C_J
\end{equation}
in the lagrangian (\ref{lagdh}) giving a gauge-invariant lagrangian. 
This gives a term linear in $C$
of the form
$C_I\we *\cj ^I$ where
$\cj ^I = J^I + *\cp ^I$, so that it differs from
$J$ by the identically conserved topological current $j^I=*\cp^I$.
The term (\ref{cj}) is then obtained by further adding a term 
\begin{equation}
\label{}
C_I\we*j ^I =
C_I\we \cp ^I
\end{equation}
 to the minimally-coupled action. However this term is not gauge-invariant and does not have a gauge-invariant completion. This is a case in which one of the obstructions to gauging of \cite{Hull:1989jk},\cite{Hull:1990ms}
  is present, and gauging is not possible.\footnote{In the terminology of \cite{Hull:1989jk},\cite{Hull:1990ms}, one is gauging the isometries generated by   $2d$ Killing vectors $k_I$ and the contraction of $ H$ with $k_I$ is
 $\i_I H =d v_I$, where $v$ is determined up to exact terms.  Choosing $v_I = \i_I b$ and using the formulae of 
 \cite{Hull:1989jk},\cite{Hull:1990ms}
 gives the gauging by minimal coupling. However, to obtain the coupling of the gauge field $C$ to $J$ instead of $\cj $ requires replacing $v$ with $v'_I=v_I+ L_{IJ}d\cx ^J$, but now
 $\i_I v'_J= L_{IJ}$ and the fact that these constants are non-zero implies that 
there is a local obstruction to gauging \cite{Hull:1989jk},\cite{Hull:1990ms}. 
However, while $v'$ is a well-defined 1-form and $J$ is a well-defined current, $v$ and $\cj $ are only locally defined, so that  the minimally-coupled action is not well-defined and  there is a topological obstruction to the gauging.
There is then an obstruction to gauging: $v$ is not globally defined, while $v'$ gives a non-zero 
$\i_{(I }v'_{J)}$ and there is no $v$ that overcomes both obstacles.}
 If one ignores global issues and gauges the symmetry generated by $\cj _I$ in (\ref{lagdh}) to obtain a local lagrangian, 
there is  a term quadratic in the gauge fields involving  $C_-^a, C_+^{a'}$  as well as
 $C_+^a, C_-^{a'}$. As this is the gauging of the symmetry (\ref{varx}), this leads to the elimination
 of all the   $\cx ^I$, leaving a sigma-model with fields $Y$  on the base space $N$.
 Thus in this case, there is an obstruction to gauging with the currents $J^I$, so that the linear term (\ref{cj}) is not obtained, and if one gauges with the currents $\cj _I$, then all of the $\cx$ are eliminated.
 
 More interesting is the gauging of $J^i$. This takes the same form as (\ref{cj}) at the linearised level, but
 the quadratic term in the gauge fields just involves  $C_+^a, C_-^{a'}$, corresponding to gauging a diagonal subgroup of the gauge group for $J^I$.
 The gauged lagrangian is $\cl _d + \cl _g+\cl _{top}$ where $\cl_d$ is the original lagrangian (\ref{lagdh}),
  $\cl _{top}$ is a topological term of the form (\ref{topo})
 and 
 \begin{equation}
\label{ksjdhf}
 \cl_g =\frac 12  C_i\we *J^i +  \frac 14 \cm ^{ij} C_i \we * C_j
\end{equation}
where
\begin{equation}
\label{}
 \cm ^{ij}=  \P^i{}_{I}  \P^j{}_{J} (\cm ^{-1})^{IJ }=
   \P^i{}_{I}  \P^j{}_{J} (L^{-1})^{IK}\cm _{KL } (L^{-1})^{LJ}
\end{equation}

This gauged lagrangian can be derived as follows.
Given a polarisation
with
\begin{equation}
  \cp =\left( \begin{array}{c} \P ^i {}_I \cp ^I
    \\  \Tilde \P _{iI} \cp ^{I}  \end{array}\right)=
\left( \begin{array}{c} P  ^i  \\ Q _i  \end{array}\right), \qquad  \hat  \cp =
\left( \begin{array}{c} \hat P  ^i  \\ \hat Q _i  \end{array}\right)
=
\left( \begin{array}{c} P  ^i +A^i \\ Q _i+\ti A_i  \end{array}\right)
\label{abc}  \end{equation} 
the lagrangian (\ref{lagdh}) can be written as
 \begin{equation}
\cl _d= \frac 1 4
G_{ij}
   \, \hat  P ^i \we *\hat  P^j + \frac 1 4
G^{ij}
   \, ( \hat  Q _i  -B_{ik} \hat P^k)  \we *( \hat  Q _j -B_{jl} \hat P^l)   
    - \half (P^i\we \ti A_i  +Q_i\we A^i) +\cl (Y) \label{lagp}  \end{equation}
The lagrangian (\ref{lagdh}) is a sigma-model on $\hat M$ and 
the symmetry being gauged is 
(\ref{varxp}), which can be viewed as an
anti-diagonal subgroup of (\ref{varx}).
Again, the first step is minimal coupling,
corresponding to making the replacement
\begin{equation}
\label{}
\cp ^I \to \cp ^I+ C_i \P ^i {}_J L^{IJ}
\end{equation}
in (\ref{lagdh}) or equivalently
 to making the replacement
\begin{equation}
\label{}
Q_i \to Q_i +C_i
\end{equation}
in (\ref{lagp}), giving a gauge-invariant lagrangian.
This has a linear coupling 
\begin{equation}
\label{}
\ha C_i \we * \cj ^i , \qquad \cj ^i = \P ^i_J j^J= J^i - \P ^i_I d\cx ^I = J^i -P^i
\end{equation}
to the Noether current $\cj$, so that adding the term 
\begin{equation}
\label{rthyfg}
\ha C_i\we P^i
\end{equation}
coupling   the gauge field  $C$ to the topological current $j^i=*P^i$
  gives the linear coupling (\ref{cji}).
In this case, the term (\ref{rthyfg}) is gauge invariant up to a surface term, so that there is no local obstruction to the gauging.\footnote{In this case, the potential obstruction to gauging is $\i ^iv ^j =  \P^i{}_{I}  \P^j{}_{J}  L^{IJ}$ and this vanishes identically.}
However, this term is not invariant under large gauge transformations.
An action invariant under large gauge transformations is given by adding the term
\begin{equation}
\label{derii}
\cl _{top}=\ha  d\ti X_i\we d X^i
\end{equation}
which when added to (\ref{rthyfg}) gives the  term
\begin{equation}
\label{rthyfg}
\ha ( d\ti X_i+C_i)\we P^i
\end{equation}
which is fully gauge-invariant under large gauge transformations.
The term (\ref{derii}) corresponds to adding the topological  term (\ref{topo})
to the classical lagrangian, with
$\W_{IJ} = \ti \P_{i[I}\P^i{}_{J]}$.

Defining
\begin{equation}
\label{}
D_i = C_i + \hat Q_i - G_{ij} * \hat P^j - B_{ij} \hat P^j
\end{equation}
the resulting lagrangian can be rewritten as
\begin{equation}
\label{}
L =  \frac 1 2
G_{ij}
   \, \hat  P ^i \we *\hat  P^j + \frac 1 2
B_{ij}
   \, \hat  P ^i \we \hat  P^j -  \hat P^i \we A_i +   
   L'
  \end{equation}
  where
  \begin{equation}
\label{}
L'= 
    \frac 1 4 
G^{ij} D_i \we *D_j  + \cl (Y) +A^i\we A_i
\end{equation}
consists of an algebraic term for the $D_i$, which are then  non-dynamical auxiliary fields
and  a term $\cl ' (Y)=\cl (Y)
+   A^i\we A_i
$ dependent only on $Y$ and given by
(\ref{lagyp}).
 In general coordinates, 
\begin{equation}
\label{}
L =  \frac 1 2
G_{ij} \x ^i _\m \x^j _\n  d\f ^\m \we * d\f^\n + \left( \frac 1 2
B_{ij} \x ^i _\m \x^j _\n  - \x ^i _\m \ti A_{i\n}\right)  d\f ^\m \we  d\f^\n  +L'
\end{equation}

Then, as was to be expected, the resulting theory is independent of $\ti X$.
Using (\ref{lagy}),(\ref{lagyp}) it
  is precisely the original theory
(\ref{lungage}) with metric $g$ given by (\ref{gis}) and $b$-field given by (\ref{bisoo}), plus  the auxiliary field term $D^2$.
The invariance under large gauge transformations means that $\ti X_i$ can be completely gauged away, including winding modes, and this is reflected in the fact that the theory is independent of $\ti X$ after integrating out the gauge fields.

The   term (\ref{derii}) is a topological term depending only on the winding numbers $n^i, \ti n_i$ of $X^i, \ti X_i$ around homology cycles in the world-sheet, so that it does not affect the classical theory.
The 
periodicities of $X, \ti X$ are   $2\p R^i$, $2\p \ti R_i$ with $\ti R_i = \a ' /R^i$
so that the $T^d$ parameterised by the $\ti X_i$ is dual to the one
 parameterised by the $  X^i$ \cite{Giveon:1991jj},\cite{Alvarez:1993qi},\cite{Hull:2006qs}.
 Then the term in the action  $S=(2\p \a ')^{-1}  \int \cl _{top}$
   is a sum of terms of the form
 $\p n^i \ti n_i $ (where  $n^i,  \ti n_i$ are winding numbers for a conjugate pair of cycles, and there is a sum over 1-cycles)
   and so contributes signs
 $e^{i\p n^i \ti n_i} = \pm 1$ to the functional integral given as a sum over winding numbers.
 A similar term arose in  \cite{Giveon:1991jj}.
 Note that changing 
the polarisation can change the sign of (\ref{derii}), but this leaves   $e^{i\p n^i \ti n_i} $ unchanged, so does not change the quantum theory. For example, changing from the $X^i$ polarisation to the $\ti X_i$ polarisation changes (\ref{derii})
 by a factor of $(-1)^d$.

Thus the gauging gives back the original sigma-model (\ref{lungage}). 
 It can also be viewed as imposing the constraint $J=0$.
 For example, choosing the gauge $C_-=0$, $C_+$ becomes a lagrange multiplier imposing
 $J_-=0$. Then the BRST constraints imply  that $J_+$  annihilates physical states, so that
 in this way the full constraint $J_\pm =0$ is achieved.

Thus given a polarisation, the constraint (\ref{scons}) can be realised  by
gauging the symmetry associated with the currents $\P J$, giving the conventional sigma-model (\ref{lungage}).
Different choices of polarisation give rise to different sigma-models and 
in each of these,  half of the coordinates $\cx$ are gauged away.
Different choices of polarisation  select a different half of the coordinates $\cx$ and are related by $O(d,d;\Z)$, and the different sigma-models obtained  are all related by T-duality.
For example, given a split $\cx \to (X^i, \ti X_i)$, choosing the polarisation as above gauges shifts in the $\ti X_i$, giving a sigma-model with coordinates $(Y, X)$, while choosing the opposite polarisation
gauges shifts in the $ X^i$, giving the dual sigma-model with coordinates $(Y, \ti X)$ (corresponding to T-dualising   all $d$ circles).

\section{T-Folds}\label{T-Folds}

A T-fold is constructed from patches in each of which there is a conventional string background, but the patching conditions involve T-dualities, and in general lead to a non-geometric background.
Let $\{U_\a\} $ be an open cover of the base $N$, $N= \cup _\a
U_\a$.\footnote{In this section $\a,\b$ will label coordinate patches and not
world-sheet coordinates.}
Then the T-fold is constructed from patches 
$U'_\a=U_\a \times T^d$, and in each such patch there is a metric $g_\a$ of the form (\ref{gis})
and a 2-form $b_\a$ of the form (\ref{bisoo}).
The metric $\bar g_\a $ and 2-form $\bar b_\a$ on $U_\a$ are patched together
in $U_\a\cap U_\b$ using diffeomorphisms and $b$-field gauge transformations in the usual way.
The remaining data specifying the geometry  consists 
of the moduli $E^\a_{ij}=G^\a_{ij}+B^\a_{ij}$ and   the $U(1)^{2d}$ connections
$A_\a, \ti A_\a$.
Over overlaps  $U_\a\cap U_\b$, these are patched together using transition functions
 in $ O(d,d;\Z)\ltimes U(1)^{2d}$, where $O(d,d;\Z)$ acts through 
 (\ref{atranss}),(\ref{tetrans}) and the $U(1)^{2d}$ acts through gauge transformations
 \begin{equation}
\label{}
\d \ca ^I= d \L ^I, \qquad \cx ^I=-\L^I
\end{equation}
This is a geometric background if the 
structure group is in the geometric subgroup $\G (d,\Z)  \ltimes U(1)^{2d}$ where
 $\G (d,\Z)  =GL(d,\Z)\ltimes \Z ^{d(d-1)/2}$ is the group of large torus diffeomorphisms and integral shifts of $B_{ij}$. Otherwise, it is a T-fold \cite{Hull:2004in}.
 
 Over each patch $U_\a$ one can instead consider a patch $U_\a \times T^{2d}$ with doubled fibre.
 As $O(d,d;\Z)\ltimes U(1)^{2d}$ acts geometrically on $T^{2d}$, with
 $O(d,d;\Z)$ acting as a subgroup of the large diffeomorphisms of $T^{2d}$, the T-fold transition functions
  in $  O(d,d;\Z)\ltimes U(1)^{2d}$ can be used for the patches
  $U_\a \times T^{2d}$ to construct a manifold $\hat M$ as a $T^{2d}$ bundle over $N$, with connection
  $\ca$ \cite{Hull:2004in},\cite{Hull:2006qs}. 
  In each patch one introduces a constant metric $L_\a$ of split signature $(d,d)$ of the form (\ref{liss})
  and a positive definite metric  $\cm _\a$ satisfying (\ref{hcons}).
  The fibre metrics $\cm _\a$ in each patch 
  transform covariantly under $O(d,d)$ (\ref{gtrans}) and so have the transition functions
  \begin{equation}
\label{}
\cm _\a =(h_{\a\b})^t \cm _\b h_{\a\b}
\end{equation}
Similar transition functions for $L$  are consistent with a constant   $L_\a = L_\b$ as the transition functions   in $O(d,d;\Z)$   preserve $L$.
    
Then for each patch, there is a doubled lagrangian
$\cl _\a$ given by (\ref{lagdh}), and in overlaps
$\cl _\a=\cl _\b $
so there is a well-defined action, which is a sigma-model with target space $\hat M$.
The constraint (\ref{scons}) is $O(d,d;\Z)$ covariant, and so is a well-defined geometric condition for the 
sigma-model on $\hat M$.

One way of imposing this constraint is to choose a polarisation and gauge, as was shown in the last section.
Consider first the case in which there are only $O(d,d;\Z)$ transition  functions, so
\begin{equation}
\label{dsasd}
\cx_\a ^I= (h^{-1}_{\a\b})^I{}_J \cx_\b ^J
\end{equation}
In each patch $U'_\a=U_\a \times T^d$, there is a choice of polarisation 
specified by projectors $\P_\a, \ti \P_\a$, which can be combined into
 a matrix 
$(\qwe _\a)^{\hat I}{}_J$, as in section \ref{polar}. This defines a splitting of the coordinates
$\cx^I_\a$ into \lq physical' coordinates $X_\a^i$ and dual coordinates
$\ti X_{\a i}$
\begin{equation}
    \bx ^{\hat I} _\a=(\qwe _\a)^{\hat I}{}_J \cx^J_\a=
    \left( \begin{array}{c} ( \P _\a)^i {}_I \cx ^I_\a
    \\  (\Tilde \P _\a)_{ iI} \cx ^{I}_\a  \end{array}\right) \label{polara}  \end{equation} 
where
\begin{equation}
    \bx ^{\hat I} _\a\equiv
    \left( \begin{array}{c} X  ^i  _\a\\ \ti X _{\a i}  \end{array}\right),
\label{polara}  \end{equation}

An active T-duality transformation transforms $\cx$ but leaves $\qwe$ invariant.
Then the transition functions (\ref{dsasd})
will give an active T-duality transformation if
 the polarisation projector  is constant, so that it is independent of the choice of patch
\begin{equation}
\label{}
\qwe_\a = \qwe_\b
\end{equation}
Then in the overlap $U'_\a \cap U'_\b$, the coordinates
$  \bx ^{\hat I} _\a$ are given by
\begin{equation}
\label{}
\bx_\a = \qwe _\a \cx_\a= \qwe _\b  h_{\a\b}^{-1} \cx_\b
\end{equation}
The term $\qwe _\b  h_{\a\b}^{-1} \cx_\b$ is regarded
as arising from   transition functions that are an active T-duality transforming   $\cx$, with
$\qwe_\a = \qwe _\b$, $\cx_\a =  h_{\a\b}^{-1} \cx_\b$.
The same ${\bf X}$ could instead be regarded as 
arising  from     a  passive T-duality acting on the polarisation
with $\qwe_\a = \qwe _\b h_{\a\b}^{-1}  $, but not on the coordinates, $\cx_\a =  \cx_\b$; in this section, the active viewpoint will be adopted,   so that $\qwe_\a = \qwe$ is independent of the patch.

Then
\begin{equation}
\label{dfgsdfg}
\bx _\a ^{\hat I}= (\hat h _{\a\b}^{-1})^{\hat I}{}_{\hat J}\bx _\b ^{\hat J}
\end{equation}
where
\begin{equation}
\label{}
\hat h _{\a\b}= \qwe  h _{\a\b}\qwe ^{-1}
\end{equation}
The matrix
$\hat h _{\a\b}$ has components
\begin{equation}
\label{}
\hat h ^{\hat I}{}_{\hat J}=
\begin{pmatrix}
 \hat h     ^i{}_j& \hat h ^{ij}  \\
     \hat h _{ij} &  \hat h_i{}^j
\end{pmatrix}
\end{equation}
so that
\begin{equation}
\label{}
X^i_\a = (\hat h _{\a\b}^{-1})^i{}_j X^j_\b +(\hat h _{\a\b}^{-1})^{ ij}\ti  X_{j \b} 
\end{equation}
In each patch, the $\{ X_\a ^i\}$ are coordinates for a $T^d$ fibre, and the condition for these 
to fit together to form a $T^d$ bundle over $N$ is that
\begin{equation}
\label{dfgsdfg}
(\hat h _{\a\b}^{-1})^{ ij}=0
\end{equation}
so
\begin{equation}
\label{}
X^i_\a = (\hat h _{\a\b}^{-1})^i{}_j X^j_\b  
\end{equation}
and the $X^i_\a$ are glued to the $X^i_\b$.
The condition (\ref{dfgsdfg}) implies that the structure group is in the geometric subgroup  
$\G(d,\Z)\subset O(d,d;\Z)$, and implies that 
the   $T^d$ fibres are patched together with diffeomorphisms
 $ (\hat h _{\a\b}^{-1})^i{}_j\in GL(d,\Z)$.
Similarly, the dual tori $\ti T^d$ will   fit together to form a bundle if 
\begin{equation}
\label{dfgsdfga}
(\hat h _{\a\b}^{-1})_{ ij}=0
\end{equation}
 and the condition for there to be both a torus bundle with fibres $T^d$ and a dual bundle with fibres $\ti T^d$ is that both (\ref{dfgsdfg}) and (\ref{dfgsdfga}) hold, so that
the structure group is in $GL(d,\Z)$.

This extends to the general case
of a T-fold with structure group in $O(d,d;\Z)\ltimes U(1)^{2d}$.
In this case it is convenient to work with the
$U(1)^{2d}$-invariant 1-forms $\Xi^I _\a$ in each patch $U'_\a$, with
\begin{equation}
   {\bf \Xi} ^{\hat I} _\a=(\qwe _\a)^{\hat I}{}_J \Xi^J_\a=
    \left( \begin{array}{c}( \P _\a)^i {}_I \Xi ^I_\a
    \\ ( \Tilde \P_\a) _{ iI} \Xi ^{I}_\a  \end{array}\right),
\label{polaracvx}  \end{equation} 
where
\begin{equation}
   {\bf \Xi} ^{\hat I} _\a\equiv
   \left( \begin{array}{c} \x  ^i  _\a\\ \ti \x _{\a i}  \end{array}\right),
\label{polarax}  \end{equation} 
Then  (\ref{dfgsdfg}) is replaced with 
\begin{equation}
\label{dfgsdfgfsd}
 {\bf \Xi} _\a ^{\hat I}= (\hat h _{\a\b}^{-1})^{\hat I}{}_{\hat J} {\bf \Xi}_\b ^{\hat J}
\end{equation}
and so
\begin{equation}
\label{}
\x^i_\a = (\hat h _{\a\b}^{-1})^i{}_j \x^j_\b +(\hat h _{\a\b}^{-1})^{ ij}\ti  \x_{j \b} 
\end{equation}
The condition that there is a $T^d$ sub-bundle is that (\ref{dfgsdfg}) holds, so that the structure group is in the 
geometric group $\G(d,\Z)\ltimes U(1)^{2d}$.

The currents $J^I_\a$ defined by (\ref{jcur}) in each patch split into the currents
 $J^i_\a, \ti J_{\a i}$
using the projectors $\P_\a, \ti \P_\a$ and  these have the transition functions
\begin{equation}
\label{}
J^i_\a = (\hat h _{\a\b}^{-1})^i{}_j J^j_\b +(\hat h _{\a\b}^{-1})^{ ij}\ti J_{j \b} 
\end{equation}
Then the constraint $J^i_\a =0$ is consistent with
$J^i_\b =0$ only if (\ref{dfgsdfg}) holds, so that the structure group is in the 
geometric group $\G(d,\Z)\ltimes U(1)^{2d}$.
If this is the case, then the constraint 
(\ref{scons})
can be imposed by gauging by coupling $J^i$ to gauge fields $C_i$.
 Note that if there are non-trivial $\G(d;\Z)$ transition functions,
 then the gauge fields $C$ are not connections on a principle bundle, but instead are connections on the affine bundle given by the pull-back of $\hat M$ to the world-sheet, with transition functions in
$  \G(d;\Z)\ltimes U(1)^d$  \cite{Hull:2006qs}.
This is sufficient to give a well-defined gauged action, even though there are no globally-defined Killing vectors \cite{Hull:2006qs}.

The bundle $\hat M$ over $N$ is characterised by
the $2d$ first Chern classes, and the $O(d,d;\Z)$  monodromies round the 1-cycles of $N$.
If all monodromies are in a subgroup  ${\cal M}\subseteq O(d,d;\Z)$, then the structure group is
in ${\cal M}\ltimes U(1)^{2d}$.
The lagrangian (\ref{lagdh}) is well-defined on $\hat M$, as is the constraint (\ref{scons}).
The constraint (\ref{scons}) can be imposed by
choosing a constant polarisation projector $\P  $, with the same choice for each patch $U_\a$, $\P_\a=\P_\b$, and then gauging the current $J^i_\a=\P^i{}_I J^I_\a$ in each patch.
The gauged lagrangians only patch together to give a   well-defined
action on $\hat M$
if $
{\cal M} \subseteq \G(d;\Z)$, so that the monodromies are all in the geometric subgroup, and in this case a geometric background is obtained.
For non-geometric  T-folds with monodromies not in the geometric group, there is no globally consistent choice of a physical $T^d$ with coordinates $X^i$, and this is reflected in the fact that the 
gauged lagrangians in each $U'_\a $ do not patch together to form a well-defined classical lagrangian on
$\hat M$.
In  the general case, the best one can do is to perform a different gauging in each patch.
These do not then fit together to form a well-defined classical action.
However, the patching is with a symmetry of the quantum theory, and the corresponding quantum theories
do patch together to give a well-defined theory, as will be discussed in the next section.

\section{Quantisation}\label{Quantisation}

In this section, the quantisation of a sigma-model on a T-fold is addressed. Suppose first the world-sheet $W$ is flat.
For the conventional formulation in terms of a  sigma-model (\ref{lungage}) with coordinates $X^i,Y^m$, one can first integrate over $X$.
For a  point $Y\in N$, the $X$ are coordinates on a torus $T^d$  
and quantising the $X$ gives the the  standard torus CFT on $T^d$ with moduli
$G_{ij}(Y),B_{ij}(Y)$.
CFT's with moduli related by $O(d,d;\Z)$ transformations are equivalent, so that $O(d,d;\Z)$ is a symmetry of the CFT, and the moduli space is not the coset $O(d,d)/O(d)\times O(d)$ parameterised by $G,B$,
but is the  Narain moduli space given by the quotient of this space by the action of  $O(d,d;\Z)$.
Then the T-fold transition functions give a bundle of torus CFT's over $N$, and this is well-defined as the 
transition functions are a CFT symmetry.

The conformal  field theory on $T^d$ can also  be formulated in an $O(d,d;\Z)$ covariant way in terms of the doubled coordinate $\cx$, imposing canonical commutation relations
on $\cx$ and its conjugate momentum.
However, in this approach one must also impose the constraint (\ref{scons})
 and the issue arises as to how to impose this in the quantum theory.
 As has been seen, this can be done by choosing a polarisation and 
 gauging the action of the current $J^i = \P^i_{I} J^I$. In general there will not be a global polarisation, and one must be chosen for each patch in $N$. One  can then quantise 
  in each patch to obtain the same torus CFT  as before
  and these patch together to give the bundle of torus CFT's over $N$.  
 
 The final stage in the quantisation is    then  to integrate over the $Y$.
The quantum theory for each patch from integrating over both $Y$ and $X$ is then the quantisation 
of the gauged sigma-model on $U_\a \times T^{2d}$.
The ungauged action (\ref{lagdh})
is a sigma-model with target space $U'=U_\a \times T^{2d}$
and is renormalizable, as is the corresponding gauged model.
The quantisation in the patch involves a choice of polarisation, but different choices lead to the same quantum theory, and can be thought of as arising from T-dual versions of the same sigma-model.

The classical lagrangian (\ref{lagdh}) is globally well-defined on $\hat M$ and is duality invariant, as is the constraint 
(\ref{scons}).  The quantisation involves choosing a polarisation that selects the independent  variables to be quantised and this breaks the duality symmetry and 
in general there is no global choice of polarisation. However, the quantum theory is duality invariant, and as the patching conditions involve a quantum symmetry, then the resulting quantum theory should be well-defined. 
It would be interesting to consider other ways of handling the constraint (\ref{scons}) in the quantum theory, and to compare the results.

Finally, in each patch, it has been seen that the two theories defined by 
the conventional sigma-model (\ref{lungage}) and  by gauging the doubled sigma-model (\ref{lagdh})
 are classically equivalent, and each is quantisable, so the question arises as to whether they define the same quantum theory.
 To quantise the gauged model, one must  first gauge-fix.
 With the topological term (\ref{derii}), the gauged action is invariant under  gauge transformations, including
 large gauge transformations specified by maps from $W$ to $U(1)^{d}$ with
non-trivial monodromy around 1-cycles in $W$.  
These can be fixed by gauging $\ti X$ away completely, using the large gauge transformations to gauge away the winding modes of $\ti X$. As was seen in section \ref{Constraint}, this gives  
the conventional  lagrangian (\ref{lungage}), plus the auxiliary field term
\begin{equation}
\label{aux}
\frac  14 G^{ij} D_iD_j
\end{equation}
In addition, there is  a ghost term
$b^ic_i$ where $b^i,c_i$ are anti-commuting scalars.
The ghost integration is trivial, so the 
result is the sum of (\ref{lungage}) and (\ref{aux}), so that quantising  the doubled formalism in this way  is equivalent to the quantisation of a conventional sigma model 
 (\ref{lungage}) plus the auxiliary term (\ref{aux}). The auxiliary field term does not affect the classical dynamics, but  as the matrix  $G^{ij}$ depends on the fields $Y$, integrating out $D_i$ give a determinant that
 affects the functional measure for $Y$. It will be seen in the next section that this change in the measure can be absorbed into a shift of the dilaton, and that this is precisely what is needed to get the the correct dilaton coupling  and transformation rules for the conventional sigma-model.
 
In this way one can define a quantum field theory for any T-fold geometry.
It remains to impose the condition that
these give modular invariant  conformal field theories, and this requires imposes   \lq field equations'    restricting the allowed backgrounds. 

\section{The Dilaton Coupling}

For curved world-sheets, one can add to the doubled sigma-model action  given by the integral of (\ref{lagdh})
the Fradkin Tseytlin term
\begin{equation}
\label{FT}
S_{FT}= \int d^2 \s \sqrt {h} \f  R
\end{equation}
where $R$ is the Ricci scalar for the world-sheet metric $h_{\a\b}$, with $h=| det(h_{\a\b})|$
and $\f$ is a scalar field on $\hat M$. It will be taken to be independent of the coordinates 
$X, \ti X$ so that it is a function $\f(Y)$ on $N$.
It is then invariant under the $O(d,d;\Z)$ symmetry of the doubled action.

On gauging and eliminating the gauge fields as in section \ref{Constraint}, one must integrate over the
auxiliary fields $D_i$ with 
lagrangian 
\begin{equation}
\label{jhkk}
   \frac 1 4 
G^{ij} D_i \we *D_j 
\end{equation}
Formally this gives a   determinant involving  $\P_\s det(G_{ij}(X(\s))$. If this is calculated as in  \cite{buscher},\cite{Tseytlin:1991wr},\cite{Schwarz:1992te}, it gives a contribution to the Fradkin-Tseytlin term  at one loop corresponding to replacing
$\f$ in (\ref{FT}) with
\begin{equation}
\label{Dil}
\F=\f - \ha \log {\rm det } (G_{ij}) = \f +\ha \log  {\rm det }  ( \P \cm  \P^{t} ) 
\end{equation}
so that the sigma-model action  on $M$ is the sum of the integral of (\ref{lungage}) and the Fradkin-Tseytlin term
 \begin{equation}
\label{FTp}
S_{FT}= \int d^2 \s \sqrt {h} \F  R
\end{equation}

Under a T-duality 
\begin{equation}
\label{}
  G^{-1}=\P \cm  \P^{t} \quad  \to \quad  (G')^{-1}=\P  h^t \cm h  \P^{t}
\end{equation}
and 
\begin{equation}
\label{iuy}
\F \to \F ' = \F + \ha \log \frac {  {\rm det }  G'}{  {\rm det }  G} 
\end{equation}
In this way, the standard T-duality transformations of the dilaton $\F$ are obtained.
There are then two dilatons, related by 
(\ref{Dil}). The dilaton $\Phi$ is the familiar one coupling to the conventional sigma-model through
the term (\ref{FTp}), transforming under T-duality
as (\ref{iuy}) and appearing as a scalar in the standard space-time effective actions.
The dilaton $\f$ coupling to the doubled sigma-model through (\ref{FT})
is invariant under $O(d,d;\Z)$ and so T-duality is a symmetry of the perturabation theory 
in the coupling constant given by the expectation value of $e^{-\f}$, but not of that defined by the 
expectation value of $e^{-\F}$. The expectation value of $e^{-\f}$ is the string field theory
coupling constant of \cite{Kugo:1992md}; see e.g. \cite{Alvarez:1996vt} for further discussion.
There will be further corrections to the relation between the two dilatons arising in this way from higher loop contributions to the change in measure \cite{Schwarz:1992te}.

\section{Doubled Everything}\label{Everything}

The doubled formulation doubles the fibre coordinates $X$ but not the base coordinates $Y$.
A more democratic and covariant formulation would be  to double the $Y$ as well.
This can always be done by adding some new coordinates
$\ti Y_m$ and then gauging the shift symmetry
$\d \ti Y_m=\ti \a_m$, or more covariantly by imposing a constraint similar to (\ref{scons}) that can be imposed by such a gauging.
The  $Y^m, \ti Y_m$ are coordinates 
on some manifold $\hat N$.
If $N$ were a torus, the $\ti Y$ could be taken as coordinates on the dual torus, but for general $N$ there is no   obvious   choice of a dual  space for $N$. 
To generalise the preceeding structure it is natural to demand that 
the tangent space $T\hat N \simeq (T\oplus T^* )N $ at each point,
so that there is a natural action of $O(n,n)$ on $T\hat N $, where $n$ is the dimension of $N$.
This suggests
taking $\hat N$   to   be the cotangent bundle $T^*N$,
or a quotient of this.

For general $M$ of dimension $D$, we then
double the coordinates $\f^\m$
to obtain
\begin{equation}
\label{Fis}
\F^M =\left( \begin{array}{c} \f^\m \\ \ti \f_\m \end{array}\right)
\end{equation}
which can be coordinates on $T^*M$ or a quotient of this.  If $M$ is a $T^d$ bundle over $N$, 
then $\hat N$ can be taken to be   a $T^{2d}$ bundle over $T^*N$, which can be thought of as a quotient
of $T^*M$ in which the coordinates $\ti X$ 
(parameterising the fibres  cotangent to $T^d$)
are periodically identified.
(In this section, $\F,\f$ are coordinates, not   dilatons.)
For the sigma model (\ref{lungage}), we introduce a
constant $O(D,D)$ invariant metric
$L_{MN}$ and
a generalised metric $\cg _{MN}$ 
satisfying
\begin{equation}
\label{s22}
\cs ^2 =\1
\end{equation}
where
\begin{equation}
\label{}
\cs = L^{-1} \cg
\end{equation}
The doubled sigma model corresponding to   (\ref{lagdh})
is then
\begin{equation}
\cl  = \frac 1 4
 \cg_{MN} \,    \cp ^M \we *  \cp ^N 
  \label{lagD}  \end{equation}
where
\begin{equation}
\label{}
\cp _\a ^M= \pa _\a \F^ M
\end{equation}
This is subject to the constraint
\begin{equation}
\label{sconD}
 \cp = \cs * \cp  
\end{equation}
(As there are no undoubled coordinates, there is no connection
$\ca$.)

The constraint (\ref{sconD}) can now be handled as in section \ref{Constraint}.
There is a natural polarisation in which
the coordinates $\f^\m$ of $M$ are selected, using a projector $\P^\m{}_M$, as the real coordinates and the coordinates of the
cotangent fibres
$\ti \f_\m$ are taken as auxiliary.
In this polarisation in which $\F$ is given in terms of $\f, \ti \f$ by
(\ref{Fis}), then  (\ref{s22}) implies $\cg$ is of the form
\begin{equation}
\cg= \left(
\begin{array}{cc}
g-bg^{-1}b & bg^{-1} \\
-g^{-1}b   & g^{-1}
\end{array}
\right).
\label{giis}  \end{equation}
for some symmetric $g_{\m\n}$
and anti-symmetric $b_{\m\n}$.
The constraint
(\ref{sconD}) is equivalent to $J^\m=0$ where
\begin{equation}
\label{}
J^\m =  \P ^\m{}_{ M} J^M, \qquad J^M = (\cs \cp - * \cp)^M
\end{equation}
The constraint $J^\m=0$ can be imposed by coupling to gauge fields as in section \ref{Constraint}, which involves
 gauging the shift symmetry  $\d \ti \f =\ti  \a$
generated by    $J^\m$, and eliminating the gauge fields and the coordinates $\ti \f$ gives precisely the original lagrangian (\ref{lungage}) (plus a topological term), by a similar argument to that given in section \ref{Constraint}.
Alternatively, if $M$ is a $T^d$ bundle over $N$, 
then $J^\m$ decomposes into $J^i, J^m$ and
one can first impose the constraint 
$J^m=0$ by coupling to gauge fields, and so
gauge the shift symmetry  $\d \ti Y= \b$ generated by $J^m$.
This eliminates the $\ti Y$ 
and   the doubled formalism lagrangian  $\cl (X,\ti X,Y) $ (\ref{lagdh})  is recovered, with the remaining constraint $J^i=0$.

For general  $T^*M$, the generalised metric
$\cg_{MN}(\f)$ depends only on the $\f$, not the $\ti \f$, so that $g(\f),b(\f)$ given by (\ref{giis})
are defined  on   $M$.
Suppose that in a patch of $M$ there are $d$ commuting Killing vectors, so that one can choose
adapted coordinates
$\f^\m= (Y^m, X^i) $ so that the Killing vectors are
$\pa/\pa X^i$ ($i=1,...,d$); at this stage, no assumptions are made about whether or not the
 $
X^i$ are  periodic.
Then the lagrangian is invariant under
shifts of $X^i,\ti X_i, \ti Y_m$
and under 
$GL(2d+n,\R)$ acting as a linear transformation on the coordinates
$X^i,\ti X_i, \ti Y_m$ (with $n=D-d$) and on $\cg$ by transformations similar to (\ref{gtrans}),(\ref{xtrans}). 
(Linear transformations involving the $Y^m$ will not be a symmetry in general if $\cg$ depends non-trivially on the $Y^m$.)
This is broken
to $O(d+n,d)$ by the constraint (\ref{sconD}), and
if $d' \le d$ of the $X,\ti X$ are periodic, then the boundary conditions   further  break the symmetry to the group
$O(d',d';\Z)\times O(d-d', d-d'+n)$.
The discrete subgroup $O(d',d';\Z)$ is a gauge symmetry of the quantum theory (provided the
 $2d'$ periodic coordinates  have the correct periodicities) with
sigma-models related by the action of $O(d',d';\Z)$ giving equivalent quantum theories.
As before, this can be thought of as changing the polarisation, so that it changes
the $d'$-dimensional subset of the $2d'$ periodic coordinates that are to be physical.
(Changing the polarisation for the non-periodic directions is not in general a gauge symmetry.)

\section{Supersymmetry}

As stated in \cite{Hull:2004in}, the supersymmetrisation  of the doubled formalism is straightforward: the sigma-model
(\ref{lagdh}) is replaced by a supersymmetric one. (The supersymmetric model was also discussed in   \cite{Hackett-Jones}.)
The N=1 supersymmetric generalisation of (\ref{lungage})
in (1,1) superspace is \cite{Gates:1984nk}
\begin{equation}
S= \frac 1 2
\int d^2\s d^2 \th \,  \left(
\gmn C^{rs} +  \bmn  \g^{rs}  
\right) D_r \f^\m D_s\f^\n
 \label{sus}  \end{equation}
where
$\f^\m(\s,\th)$ is a superfield on the superspace world-sheet with coordinates
$\s^\a, \th ^r$ where $\th^r$  are real anti-commuting coordinates transforming as a world-sheet spinor, 
 $r=1,2$ is a world-sheet spinor index,
and  $D_r$ are the usual supercovariant derivatives.
Here $C^{rs}=\e ^{rs}$ is the charge conjugation matrix and
$\g^{rs}=C^{rt} (\g_3) _t{}^s =\g^{sr}
$ where $ (\g_3) _t{}^s$ is the chirality operator satisfying
$ (\g_3)^2=1$.
The N=1 supersymmetric generalisation of (\ref{lagdh})
in (1,1) superspace is the superspace lagrangian
\begin{equation}
\cl_s=  \frac 1 4
  \cm_{IJ} \,  C^{rs}\hat  \cp ^{I}_r  \hat  \cp ^J _s- \half \g^{rs} L_{IJ} \cp ^I_r
   \ca ^J _s +\cl (Y) \label{lagds}  \end{equation}
 where
 $\cx(\s,\th), Y(\s,\th)$ are now superfields,
 \begin{equation}
\label{}
 \cp ^{I}_r = D_r \cx ^I, \qquad \hat  \cp ^{I}_r  = \cp ^{I}_r + \ca ^I_m D_r Y^m
\end{equation}
The superspace versions  of (\ref{lagyp}),(\ref{lagy}) are
\begin{equation}
\label{lagyps}
\cl (Y)
=\cl '(Y)
- \g^{rs}  A^i_r   A_{si}
\end{equation}
and
  \begin{equation}
\label{lagys}
\cl ' (Y)= \ha 
\left( \bar g_{mn} C^{rs}+ \bar b_{mn}  \g^{rs} 
\right) D_rY^m D_sY^n
\end{equation}
The supersymmetric version of the topological term
(\ref{topo}) is
\begin{equation}
\label{topos}
\cl_{top}= \ha \W _{IJ}  \g^{rs} 
\cp ^{I}_r \cp ^{J}_s 
\end{equation}
and the component  expansion of this gives the  topological term
(\ref{topo}) plus the total derivative of a  fermion bilinear.

The constraint (\ref{scons}) becomes
\begin{equation}
\label{sconss}
\hat \cp = S\g _3 \hat \cp  
\end{equation}
The component expansion gives fermionic bilinear contributions to the constraint (\ref{scons}), and
a constraint on the world-sheet fermions $\psi^I$ which reduces
to $\psi= S\g _3 \psi$ in the free case, so that $\psi ^a$ is a left-handed chiral spinor 
and $\psi ^{a'}$ is a right-handed one.

As in the bosonic case, this can be imposed by choosing a polarisation and gauging as in section \ref{Constraint},
coupling to a superspace gauge field $\G_{ri}$. The superspace current $J^i_r$
corresponding to  (\ref{jcur})
is
\begin{equation}
\label{}
J^i_r = \P^i{}_I J^I_r
\end{equation}
where
\begin{equation}
\label{jadsdjasl}
J^I_r=L^{IJ}J_{rJ}=S^I{}_J \hat P^J_r- (\g_3)_r{}^s\hat P^I_s
\end{equation}
The supersymmetric gauging is given by adding to  (\ref{lagds}) the supersymmetric generalisation of 
(\ref{ksjdhf})
given by
 \begin{equation}
\label{}
\cl_g= \frac 12   C^{rs} \G_{ri} J^i _s+  \frac 14 \cm ^{ij}  C^{rs} \G_{ri}  \G_{sj}
\end{equation}
Then eliminating the gauge field and
$\ti X_i$ as in section \ref{Constraint}, one  recovers the lagrangian (\ref{sus}), giving the local equivalence of the formalisms.
The discussion of quantisation and global structure extend straightforwardly to the supersymmetric case.

The formulation of section \ref{Everything} also generalises straightforwardly to superspace
giving the superspace lagrangian
\begin{equation}
\cl  = \frac 1 4 
 \cg_{MN} C^{rs} \,    \cp ^M_r   \cp ^N _s
  \label{lagDasd}  \end{equation}
  subject to the constraint
\begin{equation}
\label{sconDas}
 \cp = \cs \g_3 \cp  
\end{equation}

\end{document}